%% file: jspaper.tex
\documentclass[smallextended]{svjour3}
\usepackage{xspace}
\usepackage{paralist}
\usepackage{graphicx}
\usepackage[utf8]{inputenc}
\usepackage{listings}
\usepackage{color}
\usepackage{marvosym}
\usepackage{mathtools}
\usepackage{amsfonts}

\usepackage{fancybox}
\usepackage{booktabs}
\usepackage{algorithmicx}
\usepackage{algpseudocode}
\makeatletter
\renewcommand{\ALG@beginalgorithmic}{\small}
\makeatother
\usepackage{multirow}
\usepackage[binary-units]{siunitx}
\usepackage{url}

\newcommand{\ff}{\textsc{FastFlow}\xspace}
\newcommand{\fftitle}{FastFlow\xspace}
\newcommand{\srl}[0]{\emph{Loop-of-stencil-reduce}\xspace}
\newcommand{\srli}[0]{\emph{Loop-of-stencil-reduce-i}\xspace}

\newcommand{\code}[1]{#1}
\newcommand{\tsize}{\scriptsize}

\lstnewenvironment{Bench}[2]
{\lstset{ %
showlines=true,
xleftmargin=5ex,
firstnumber=1,
language=C++,                
basicstyle=\small,       
numbers=left,                   
numberstyle=\small,      
stepnumber=1,                   
numbersep=3pt,                  
showspaces=false,               
showstringspaces=false,         
showtabs=false,                 
tabsize=2,                      
captionpos=b,                   
breaklines=true,                
breakatwhitespace=false,        
escapeinside={\%*}{*)},          
mathescape=true,
columns=flexible,
morekeywords={main,add_workers,add_emitter,wrap_around,
run_and_wait_end,ff_send_out,stencil,after,before,prepare,reduce,function,read},
#1
}}{}

\begin{document}
\title{A Parallel Pattern for Iterative Stencil + Reduce }

\author{M. Aldinucci \and M.Danelutto \and M. Drocco \and P. Kilpatrick \and C. Misale \and G. Peretti Pezzi \and M. Torquati}
\authorrunning{Aldinucci et. al.}

\institute{M. Danelutto and M. Torquati \at
              Dep. of Computer Science, University of Pisa, Italy \\
              \email{\{marcod, torquati\}@di.unipi.it}
           \and
           M. Aldinucci, M. Drocco and C. Misale \at
              Dep. of Computer Science, University of Turin, Italy \\
              \email{\{aldinuc, drocco, misale\}@di.unito.it}
           \and
           P. Kilpatrick \at
           Dep. of Computer Science, Queen's University Belfast, UK \\
           \email{p.kilpatrick@qub.ac.uk}
           \and
           G. Peretti Pezzi \at
              Swiss National Supercomputing Centre, Switzerland \\
              \email{gppezzi@gmail.com}
}

\date{Received: date / Accepted: date}

\maketitle

\begin{abstract} 
We advocate the \srl pattern as a means of simplifying the
implementation of data parallel programs on heterogeneous multi-core platforms.
\srl is general enough to subsume \emph{map}, \emph{reduce}, \emph{map-reduce}, 
\emph{stencil}, \emph{stencil-reduce}, and, crucially, their usage in
a loop in both data parallel and streaming applications, or a combination of
both.
The pattern makes it possible to deploy a single stencil computation kernel on different GPUs.
We discuss the implementation of \srl in \ff, a framework for implementation of
applications based on parallel patterns. Experiments are presented to illustrate
the use of \srl in developing data-parallel kernels running on heterogeneous systems. 
\end{abstract}

\section{Introduction}\label{sec:Introduction}

Data parallelism has played a paramount role in application design
from the dawn of parallel computing. \emph{Stencil}
kernels are the class of (usually iterative) data parallel kernels which update
array elements according to some fixed access pattern.
The stencil paradigm naturally models a wide class of algorithms
(e.g. convolutions, cellular automata, simulations) and it typically 
requires only a fixed-size and compact data exchange among processing
elements, which might follow a weakly ordered execution 
model. The stencil paradigm does not exhibit true data dependencies within a single iteration.  
This ensures  efficiency and scalability on a wide range of
platforms ranging from GPUs to clusters.
GPUs are widely perceived as data-parallel computing systems~\cite{owens:sc07:dp:gpu} 
so that GPU kernels are typically designed to employ the \emph{map-reduce} parallel paradigm.
The \emph{reduce} part is typically realised as a sequence of partial GPU-side reduces, 
followed by a global host-side reduce. 
Thanks to GPUs' globally shared memory, a \emph{map} computation can
implement a stencil as a data overlay with non-empty intersection,
provided they are accessed in read-only fashion to enforce deterministic behaviour.   
Often, this kind of kernel is iteratively called in host code in a loop body up to
a convergence criterion.

Data parallelism has been provided to application programmers by way of various code artefacts
(constructs, from now on) in both  shared-memory and message-passing programming models (e.g. compiler
directives, skeleton frameworks, pattern libraries). Its 
implementation is well understood for a broad class of
platforms, including GPUs (see Sec. \ref{sec:RW}). In this setting,
the possibility to compose constructs certainly enhances expressiveness
but also the complexity of the run-time system.

We advocate composition beyond the class of data parallel constructs. 
We envisage parallelism exploited according to the \emph{two tier} model~\cite{assist:cunhabook:05}: 
\emph{stream} and \emph{data} parallel. Constructs in each tier can be
composed and data parallel constructs can be nested within stream
parallel ones. 
The proposed approach distinguishes itself from nesting of \emph{task} and \emph{data} parallelism, which
has been proposed (with various degrees of integration) as a way to integrate
different platforms:
examples include MPI+OpenMP, OmpSs+SkePU, MPI+CUDA. 
These approaches naturally target a two-tier platform
(e.g. cluster of multicores), whereas a composition of patterns can be
mapped onto multiple hardware tiers, each one exhibiting
a different synchronisation latency. Whatever an extreme scale
platform will be, it will be built across multiple tiers. 

In this setting, we proposed the \srl pattern~\cite{opencl:ff:ispa:15} as an abstraction for tackling the complexity of 
implementing iterative data computations on heterogeneous platforms.
The \srl is designed as a \ff~\cite{fastflow:parco:09,tutorial:ff:15} pattern, which can be nested in other stream parallel patterns, 
such as \emph{farm} and \emph{pipeline}, and implemented in C++ and OpenCL.
We advocate it as a comprehensive pattern for programming GPUs
in a way that is general enough to express \emph{map}, \emph{reduce}, \emph{map-reduce}, \emph{stencil},
\emph{stencil-reduce} computations and, most significantly, their usage in a loop.

The \srl simplifies GPU exploitation by taking care of a number of low-level issues, such as:
device detection, device memory allocation, host-to-device (H2D) and
device-to-host (D2H) memory copy and synchronisation, reduce algorithm
implementation, management of persistent global memory in the
device across successive iterations, and enforcing data race avoidance
due to stencil data access in iterative computations.
%
Finally, it can transparently exploit multiple GPUs on the same platform.


%
While this paper builds on previous results \cite{ff:denoiser:ijhpca:15,opencl:ff:ispa:15}, 
it advances  them in several directions.
%
The \srl pattern is an evolution of the
\emph{stencil-reduce} pattern
\cite{ff:denoiser:ijhpca:15} and it has been refined to
explicitly include the iterative behaviour and the optimisations
enabled by the awareness of the iterative computation and the possible
nesting into a streaming network.
Such optimisations are related to GPU persistent
global memory usage, stencil and reduce pipelining, and asynchronous
D2H/H2D memory copies. The \srl has been uniformly implemented in OpenCL and
  CUDA, whereas \emph{stencil-reduce} was dependent on CUDA-specific
  features not supported in OpenCL, such as Unified 
  Memory. Also, locally-synchronous computations (by way of halo-swap) across multiple GPUs have been
introduced, whereas in previous works use of 
  multiple GPUs was possible only on independent kernel instances.
The paper itself extends \cite{opencl:ff:ispa:15} by
introducing a formalisation of the \srl pattern, and a brand new 
experimentation plan.  Specifically, the paper extends the previous experimentation by reporting tests on 
three applications  and three different heterogeneous platforms, by also demonstrating that it is possible to derive a \srl formulation of three different applications. 
Two applications out of three exploit both stream and data parallelism. The set of
platforms includes a multiple NVidia GPU Intel box and a ``big.LITTLE'' Samsung
mobile platform with 2 different Arm multi-core CPUs and 1 Arm GPU. 

 




\input{relw}

\section{The \srl  pattern in \fftitle}
\label{sec:srl}

In this section the semantics and the \ff implementation of \srl are
introduced. The well-known Conway's Game-of-life is used as a simple but
paradigmatic example of locally synchronous data-parallel
applications (running on multiple devices).
The provided semantics of stencil computations considers only symmetric stencils with a regular topology of the neighbours.

\subsection{Semantics of the \srl pattern}
\label{sub:srl math}
We assume that $a$ is an $n$-dimensional array with dimension sizes $d_1,
\ldots, d_n$ and items of type $T$.  We define the \emph{apply-to-all} functional $\alpha(f)$ as follows:
$$
(\alpha(f):a)_{i_1, \ldots, i_n} = f(a_{i_1, \ldots, i_n})
$$ 
where ``:'' denotes the function application, $f$ has type $T \to T'$ and $\alpha(f):a$ is an array of the same size as $a$ and items of type $T'$.
We also define $/(\oplus)$ as:
$$
(/(\oplus):a)_{i_1, \ldots, i_n} = {{\bigoplus}}_{\forall i_1 \in [0,d_1-1]; \ldots; \forall i_n \in [0,d_n-1]} (a_{i_1,\ldots, i_n})
$$
where $\oplus$ is a binary and associative operation with type $T
\times T \to T$, and $\bigoplus_{i=\ldots} x_{i}$ ``sums'' up all the $x_i$ by means of the $\oplus$.
%
%
Then we define the generic $n$-dimensional \textit{stencil operator} $\sigma_k^n$ as follows:
\[
\left\{
\begin{array}{ll}
(\sigma_k^n : a)_{i_1, \ldots, i_n} = w_{i_1, \ldots, i_n} \in T^{(2k+1)^n}\\
(w_{i_1, \ldots, i_n})_{j_1, \ldots,j_n} = a'_{i_1-k+j_1,\ldots,i_n-k+j_n}, j_l \in [0,2k+1]
\end{array}
\right.
\]
where neighbourhoods $w_{i_1, \ldots, i_n}$ have $2k+1$ items for each dimension and $a'_{i_1, \ldots, i_n}=\bot$ if some index $i_l$ falls out of the dimension range $[0,d_l-1]$ while $a'_{i_1, \ldots, i_n}=a_{i_1, \ldots, i_n}$ otherwise.

With these definitions, we proceed to characterise the stencil parallel pattern functional semantics as\footnote{We omit the dimension $n$ in $\sigma^n_k$ here, as we assume the dimension $n$ is the same as that of the array $a$: a single dimensional array will have $n=1$, a 2D matrix $n=2$, and so on.}
${\textbf{stencil}}(\sigma_k,f) : a = \alpha(f) \circ \sigma_k : a$,
possibly computing in parallel all the
$f(w_{i_1, \ldots, i_n})$
applications. We remark that, in this formulation, $f$ takes as input a neighbourhood of type
$T^{(2k+1)^n}$.
Moreover, both $f$ and $\oplus$ should take into account the possibility that some of the input arguments are $\bot$. At this point we may formally define the \srl parallel pattern's functional semantics as follows:
  \begin{algorithmic}[1]
    \Procedure{loop-of-stencil-reduce}{($k,f,\oplus,c,a$)}
    \Repeat
    \State $a$ = $\mbox{\textbf{stencil}}(\sigma_k,f):a$
    \Until{c($/\oplus:a$)}
    \EndProcedure
  \end{algorithmic}
 
We consider this as the simplest pattern modelling iterative stencil+reduce parallel computations.
Small variants of this pattern are worth consideration, however, to
take into account slightly different computations with similar parallel
behaviour.
The first variant considered is that where the function applied
in the $\alpha(f)$ phase takes as an input the ``index'' of the
element considered (the centroid of the neighbourhood) in addition to
all the items belonging to the neighbourhood.
We call this variant \srli and it can be
simply defined by the same algorithm as that of the
\srl with minor changes to the
auxiliary functions $f$ and $\oplus$:
\begin{itemize}
\item we consider a new function $\overline{f}$ of type $(T\times \mathbb{N}^n)^{(2k+1)^n} \to T'$, thus working on neighbourhoods composed of value-index pairs;
\item a new stencil operator $\overline{\sigma}^n_k$ enriching neighbourhoods with indexes:
$$
\left\{
\begin{array}{ll}
(\overline\sigma_k^n : a)_{i_1, \ldots, i_n} = \overline w_{i_1, \ldots, i_n} \in (T\times \mathbb{N}^n)^{(2k+1)^n}\\
(\overline w_{i_1, \ldots, i_n})_{j_1, \ldots,j_n} =
\langle a'_{i_1-k+j_1,\ldots,i_n-k+j_n},
\langle
i_1-k+j_1,\ldots,i_n-k+j_n
\rangle
\rangle
\end{array}
\right.
$$
\end{itemize}
where $j_l \in [0,2k+1]$.
With such definitions the \textit{loop-of-stencil-reduce-i} is just a
\srl
with different parameters, that is
  {\small \sc
    {Loop-of-stencil-reduce}}$\left(k,\overline{f},\overline{\oplus},c,a\right)$.
\noindent
The second variant of the \srl pattern we introduce  changes slightly the way in which the termination condition is computed and used, to deal with those iterative computations where \textit{convergence} of the reduced values is of interest, rather than their absolute values.
We consider:
\begin{itemize}
\item a new function $f'$ returning also the input value:
$$
f':a_{i_1, \ldots, i_n} = \langle f : a_{i_1, \ldots, i_n}, a_{i_1, \ldots, i_n} \rangle
$$
\item $\delta$ of type $T \times T \to T$, that is applied over all
    the items resulting from the $\alpha(f)\circ\sigma_k$
    step to combine contributions of the two most recent iterations;
  \item $\oplus$ of type $T \times T \to T$, that is used to reduce
    the items computed by $\delta$ to a single value to be passed to
    termination condition $c$. 
  \end{itemize}
With these definitions, we may define the second \srl variant as follows:

\begin{algorithmic}[1]
  \Procedure{loop-of-stencil-reduce-d}{($k,f,\delta,\oplus,c,a$)}
  \Repeat
  \State $ b = $ \textbf{stencil}($\sigma_k,f'$):$a$
  \State $ d = \alpha(\delta):b$
  $\quad$ $ a = \alpha(\mbox{\it fst}):b$ \Comment{being $\mbox{\it fst} :\langle a,b \rangle
    = a$}
  \Until{c($/\oplus:d$)}
  \EndProcedure
\end{algorithmic}

It is clear that the {\small \sc loop-of-stencil-reduce-d} may be
easily extended to a {\small \sc loop-of-stencil-reduce-d-i} where the $\overline{f}$ and $\overline{\sigma}_k$ functions are used in place of $f$ and $\sigma_k$ as we did to turn the \srl into \srli.
The third and last variant we present simply consists in considering
some kind of global ``state'' variable (such as the number of
iterations) as a parameter of the termination condition:

  \begin{algorithmic}[1]
    \Procedure{loop-of-stencil-reduce-s}{($k,f,\oplus,c,a$)}
    \State $s$ = init($\ldots$);
    \Repeat
    \State $a$ = $\mbox{\textbf{stencil}}(\sigma_k,f):a$; 
    $\quad$
    $s$ = update($\ldots$);
    \Until{c($/\oplus:a$,s)}
    \EndProcedure
  \end{algorithmic}
  and again it may be included in both the {\small{\sc{-d}}} and {\small{\sc{-i}}} versions of the \srl pattern.

With a similar methodology, we may define the functional semantics of more classical data parallel patterns such as \textbf{map} and \textbf{reduce}:
the \textbf{map} pattern computes
   $\mbox{\textbf{map}}(f):a = \alpha(f):a$
  possibly carrying out all the $f(a_{i_1, \ldots, i_n})$ computations in parallel and
  the \textbf{reduce} pattern computes
  $\mbox{\textbf{reduce}}(g):a = /(g):a$
  possibly computing in parallel the different applications of $g$ at the same level of the resulting reduction tree.



We remark that, from a functional perspective, \textbf{map} and
\textbf{stencil} patterns are very similar, the only difference being
the fact that the stencil elemental function $f$ takes as input a set
of atomic elements rather than a single atomic element.  Nevertheless,
from a computational perspective the difference is substantial, since
the semantics of the map leads to \emph{in-place} implementation,
which is in general impossible for stencil.
These
parallel paradigms have been proposed as patterns  for both multi-core
and distributed platforms, GPUs,
and heterogeneous platforms \cite{ske:survey:spe10,Enmyren:2010:SMS:1863482.1863487}.
They are well-known examples of 
data-parallel patterns since, as stated above, the elemental function of a map/stencil can be
applied to each input element independently  of the others, and also 
applications of the combinator to different pairs in the reduction tree of a
reduce can be done independently, thus naturally inducing a parallel
implementation.
Finally, we remark that the basic building block of \srl (the
\textit{repeat} block at lines 2--4 of the {\sc
  loop-of-stencil-reduce} pattern above) is
 \textit{de-facto} the stencil-reduce pattern
previously presented in~\cite{ff:denoiser:ijhpca:15}.

\subsection{The \ff \srl API}
\label{sub:srl api}
At high level, \ff applications are combinations of higher-order
functions called parallel patterns \cite{fastflow:parco:09,tutorial:ff:15}. 
A \ff pattern describes the functional transformation from input to output streams.
Some special patterns, referred to as data-parallel patterns, exhibit parallelism
by applying the same function to each element of an input set.
In particular, the \srl pattern implements an instance of the semantics described in~\ref{sub:srl math}
in which the stencil-reduce computation
is iteratively applied, using the output of the stencil at the $i$-th iteration
as the input of the $(i+1)$-th stencil-reduce iteration.
Moreover, it uses the output of the reduce computation at
the $i$-th iteration, together with the iteration number, as input of the \emph{iteration condition},
which decides whether to proceed to iteration $i+1$ or stop the computation.

The \ff implementation is aimed at supporting iterative data-parallel computations both on CPU-only and CPU+GPU platforms. 
For CPU-only platforms, the implementation is written in C++ and exploits the \ff map pattern.
On the other hand, when an instance of the \srl pattern is deployed onto GPUs or
other accelerators\footnote{the current implementation does not allow
 mixing of CPU and GPUs (or other accelerators) for deploying a single \srl instance.},
the implementation relies on the OpenCL framework features.
The \ff framework provides the user with constructors for building
\srl instances, i.e. a combination of parametrisable building
blocks:
\begin{compactitem}
\item the OpenCL code of the elemental function
of the stencil;
\item the C++ and OpenCL codes of the combinator function;
\item the C++ code of the iteration condition.
\end{compactitem}
The language for the \emph{kernel} codes implementing the elemental function and the combinator
-- which constitute the business code of the application -- can be
device-specific or coded in a suitably specified C++ subset
(e.g. REPARA C++ open specification \cite{repara-D2.1}).
Functions are provided that take as input the
business code of a kernel function (elemental function or combinator) and translate it to a fully defined OpenCL kernel, which will be offloaded to target accelerator devices by the \ff runtime.
Note that, from our definition of elemental function (Sec.~\ref{sub:srl math}), it follows that the \srl programming model is data-oriented rather than
thread-oriented, since indexes refer to the input elements rather than
the work-items (i.e. threads) space,  which
is in turn the native programming model in OpenCL.

When instantiating a \srl pattern, the user may also
specify some non-functional parameters for controlling parallelism such as
the type and number of accelerator devices to be used (e.g.
number of GPUs in a multi-GPU platform) and the maximum
size of the neighbourhood accessed by the elemental function.
Note that the latter parameter can be determined by a static
analysis on the kernel code in most cases of interest,
i.e. ones exhibiting a static stencil (e.g. Game of Life~\cite{Gardner:1970:MGFa}) or
dynamic stencil with reasonable static bounds (e.g. Adaptive Median
Filter, \cite{ff:denoiser:ijhpca:15}). 

Multi-GPU environments can be exploited in two different ways,
namely either each item from the input stream is sent to a single GPU (i.e. 1:1 mode)
or a single item is sent to a $n$-GPU \srl pattern\footnote{a $n$-GPU
pattern is a pattern deployed onto $n$ GPU devices.}. The latter case yields
$n$ GPUs processing each input item in parallel. We refer to the two cases as 1:1 and 1:$n$ modes, respectively.
Although this poses some challenges at the \ff implementation level
(see Sec.~\ref{sub:srl impl}),
it requires almost negligible modifications to user code.
That is, when defining the OpenCL code of the elemental function, the
user is provided with local indexes over the index space of the
device-local sub-input~-- e.g. for accessing input data~--
along with global indexes over the index space of the whole input~--
e.g. for checking the absolute position with respect to input size.
For the case 1:$n$, the input item is split evenly for 1D array and 
by rows for 2D matrix.

Figure~\ref{fig:gol srl ff} illustrates a Game of Life implementation
on top of the \srl API in \ff. Source-to-source functions are used
to generate OpenCL kernels for both stencil elemental function
(lines 1--12) and reduce combinator (lines 14--15). The source codes are wrapped into fully defined kernels, automatically optimised by the OpenCL runtime system.. 
The user, in order to exploit
1:n parallelism, has to use local indexes \code{i\_} and \code{j\_}
to access elements of the input matrix. C++ codes for
iteration condition (\code{iterf}) and reduce combinator (\code{reducef}) are not reported, as
they are trivial single-line C++ lambdas.
The constructor (lines 17--20) builds a \srl instance by taking
the user-parametrised building blocks as input, plus the identity element
for the reduce combinator (0 for the sum) and the parameters for controlling
1:$n$ parallel behaviour, namely the number of devices to
be used over a single item (\code{NACC}) and the 2D maximum sizes of
the neighbourhood accessed by the elemental function  
(Game of Life is based on 3-by-3 neighbourhoods).
Finally, the constructor is parametrised
with a template type \code{golTask} which serves as an interface for
basic input-output between the application code and the \srl
instance. 

\begin{figure}
\begin{Bench}{}{}
std::string stencilf = ff_stencilKernel2D_OCL(
                   "unsigned char", "in", //element type and input
                   "N", "M",              //rows and columns
                   "i", "j", "i_", "j_",  //row-column global and local indexes
                   std::string("") +  
           /* begin of the OpenCL kernel code */
                   "unsigned char n_alive = 0;\n" + 
                   "n_alive += i>0 && j>0 ? in[i_-1][j_-1] : 0;\n" +
                   ... + 
                   "n_alive += i<N-1 && j<M-1 ? in[i_+1][j_+1] : 0;\n" +
                   "return (n_alive == 3 || (in[i_][j_] && n_alive == 2));"
           /* end OpenCL code */);						

std::string reducef = ff_reduceKernel_OCL(
                   "unsigned char", "x", "y", "return x + y;");

ff::ff_stencilReduceLoop2DOCL<golTask> golSRL(
                   stencilf, reducef, 0, iterf, // building blocks
                   N, N, NACC,    // matrix size and no. of accelerators
                   3, 3);         // halo size on the 2 dimensions
\end{Bench}
\caption{Implementation of Game of Life~\cite{Gardner:1970:MGFa} on top of the \srl API in \ff.\label{fig:gol srl ff}}
\end{figure}

\ff does not provide any automatic facility to convert C++ code into
OpenCL code, but 
facilitates this task via a number of features including:
\begin{compactitem}
\item Integration of the same pattern-based parallel programming model
  for both CPUs and GPUs. Parallel activities running on CPUs can be
  either coded in C++ or OpenCL.
\item Setup of the OpenCL environment.
\item Simplified data feeding to both software accelerators
  and  hardware accelerators (with asynchronous H2D and
  D2H data
  movements). 
\item Orchestration of parallel activities and synchronisations within
  kernel code (e.g. reduce tree), synchronisations among kernels
  (e.g. stencil and reduce in a loop), management of data copies
  (e.g. halo-swap buffers management).
\item Transparent usage for the user of multiple GPUs on the same platform.
\end{compactitem}

\subsection{The \ff implementation}
\label{sub:srl impl}
The iterative nature of the \srl computation
presents challenges for the management of
the GPU's global memory across multiple iterations, i.e. across
different kernel invocations.
The general schema of the \srl pattern is described in Fig.~\ref{fig:stencil-reduce}.
Its runtime is tailored to efficient loop-fashion execution. When a task\footnote{we implicitly
	define a \ff task as the computation to be performed over a single stream item by a \ff pattern.}
is scheduled
to be executed by the devices the pattern is deployed onto, the runtime takes care of allocating
on-device global memory buffers and filling them with input
data via H2D copies.
The na\"{i}ve approach for supporting iterative computations on a hardware accelerator device equipped with some global memory (e.g. GPU) would consist in putting a global synchronisation barrier
after each iteration of the stencil, reading the result of the stencil back from the device buffer (full size D2H copy),
copying back the output to the device input buffer (full size H2D copy) and proceeding to the next iteration.
\ff in turn employs \emph{device memory persistence} on the GPU across
multiple kernel invocations, by just swapping on-device buffers. In the case of a multi-device 1:$n$ deployment (Sec.~\ref{sub:srl api}),
small device-to-device copies are required after each iteration, in order to keep
halo borders aligned, since no device-to-device copy mechanism is available
(as of OpenCL 2.0 specification, device-to-device transfers).
Global memory persistence is quite common in
iterative applications because it drastically reduces the need for
H2D and D2H copies, which can severely limit the
performance. This also motivates the explicit inclusion of the iterative behaviour
in the \srl pattern design which is one of the differences with respect 
to solutions adopted in other frameworks, such as SkePU \cite{Enmyren:2010:SMS:1863482.1863487}.
As a further optimisation, \ff exploits OpenCL events to keep \srl computation as asynchronous as possible.
In particular, in the case of a multi-GPU 1:$n$ deployment, memory operations
and sub-tasks running on different GPUs at the same iteration 
are independent of each other, and so  can run in parallel.
The current implementation employs simple heuristics (basically wrappers of OpenCL routines) to determine the kernel launching parameters for controlling the layout of OpenCL threads.

\begin{figure}
\begin{Bench}{}{}
while (cond) { 
  before (...)  // [H] initialisation, possibly in parallel on CPU cores
  prepare (...) // [H+D] swap I/O buffers, set kernel args, D2D-sync overlays
  stencil<SUM_kernel,MF_kernel> (input, env)  // [D] stencil and partial reduce
  reduce op data  // [H] final reduction
  after (...)  // [H] iteration finalisation, possibly in parallel on CPU cores
}
read(output) //[H+D] D2H-copy output
\end{Bench}
\caption{\srl pattern general schema.\label{fig:stencil-reduce}}
\end{figure}

\input{exp}

\section{Conclusions}\label{sec:conl}

In this work we built upon the \srl parallel pattern~\cite{opencl:ff:ispa:15}, an evolution of the stencil-reduce pattern presented in~\cite{ff:denoiser:ijhpca:15} targeting iterative data-parallel computations on heterogeneous multi-cores.
We first provided  motivation and then gave a semantics for the pattern.
Furthermore, we showed that various iterative kernels can be easily and effectively parallelised by using the \srl on the available GPUs
by exploiting the OpenCL capabilities of the \ff parallel framework.

We have focused here on capturing stencil iteration as a pattern, and on its integration in the established \ff pattern framework. Much work has been done elsewhere  on optimisation of stencil implementations on GPUs (e.g. in \cite{Lutz:2013:PAF:2400682.2400718}) and we intend in the future to incorporate such optimisations into our \ff implementation. As a further extension, we plan to build on top of the current implementation of the \srl a domain specific language (DSL) 
specifically targeting data parallel computations in a streaming work-flow. This extension will not substitute the current interface but it will be a further layer. Thus, the current expressiveness would not be affected by the DSL.

\medskip

\noindent \textbf{Acknowledgment}
This work was supported by EU FP7 project REPARA 
(no. 609666), the EU H2020 project RePhrase (no. 644235) and by the NVidia GPU Research
Center at University of Torino.

\bibliographystyle{spmpsci}
\bibliography{biblio/skeletons_ac_grid,biblio/Biblio,biblio/UniPisaTorinoGroup,biblio/multicore,biblio/extra,biblio/repara}


\end{document}

%% file: relw.tex
\section{Related Work}\label{sec:RW}

Software engineers are often involved in solving recurring
problems. Design patterns have been introduced to provide effective
solutions to these problems. Notable examples are stream parallel
patterns, such as \emph{farm} and \emph{pipeline}, and data parallel patterns such as \emph{map}, \emph{reduce} and
\emph{stencil}.
%
%
Several parallel programming frameworks based on patterns target heterogeneous platforms. Here we consider a selection of the most well known.

In Muesli \cite{muesli-gpu} the programmer must explicitly
indicate whether GPUs are to be used for data parallel skeletons. 

StarPU \cite{Augonnet:2011:StarPU} is focused on handling 
accelerators such as GPUs. Graph tasks are scheduled by its run-time 
support on both the CPU and on various accelerators, provided the programmer 
has given a task implementation for each architecture. 
%

The SkePU programming framework \cite{Enmyren:2010:SMS:1863482.1863487} provides programmers with 
GPU implementations of several data parallel skeletons (e.g. Map, MapOverlap, MapArray,
Reduce) and relies on StarPU for the execution of stream
parallel skeletons (pipe and farm). 



In SkelCL~\cite{SkelCL2013}, a high-level skeleton library built on top of OpenCL code, container data types are used 
to automatically optimize data movement across GPUs. Recently, two new SkelCL skeletons targeting stencil computations have been introduced~\cite{breuer.2014.histencils}:
the MapOverlap skeleton for single-iteration stencil computations and the Stencil skeleton that provides more complex stencil patterns and iterative computations.

The \ff stencil operation is similar to both the Stencil skeleton in SkelCL, 
and to the SkePU overlay skeleton. 
The main difference is that they rely on specific internal data types. Furthermore, to the best of our knowledge, SkePU 
is not specifically optimised for iterative stencil computation whereas SkelCL provides iterative computations but the 
current version handles only iterative loops with a fixed number of iterations. However, they plan to allow the user to specify 
a custom function as it is currently provided in the \srl.


In this context, the \ff parallel programming environment has recently been
extended to support GPUs via CUDA \cite{ff:denoiser:ijhpca:15} and
OpenCL (as described in the present work). \ff CPU implementations of patterns
are realised via non-blocking graphs of threads connected by way of
lock-free channels~\cite{ff:acc:europar:11}, while the GPU implementation is
realised by way of the OpenCL bindings and offloading 
techniques. 
Also, different patterns can be mapped onto
different sets of cores or accelerators and so, in principle, can use the full available
power of the heterogeneous platform.

Among compiler-based approaches, we recall OpenACC and OmpSs, differing from the \ff approach since it consists of a header files library. They do not provide any stencil pattern but they focus on loop parallelism with offloading.
OpenACC~\cite{openacc} is a compiler-based, high-level, performance portable
programming model that allows programmers to create
high-level host\-+\-accelerator programs
without the need to explicitly initialise the accelerator or manage data transfers
between the host and accelerator. It is based on compiler directives, such as
pragmas, that, for instance,  allow execution of a loop on a GPU by just adding the parallel loop. It also supports multi-GPU execution.
The task-based OmpSs~\cite{ompss:12:IPDPS} extends OpenMP with directives to
support asynchronous parallelism and heterogeneity, built on top of
the Mercurium compiler and Nanos++ runtime system. Asynchronous
parallelism is enabled by the use of data-dependencies between the
different tasks of the program, and execution on multi-GPU is also
supported. 

For an extensive discussion of the state of the art on compiler and
dynamic optimisations possible on stencil computations on GPUs we refer
to \cite{Lutz:2013:PAF:2400682.2400718}.

%% file: exp.tex
\section{Experiments}
\label{sec:exp}
Here we present an assessment of the \srl \ff implementation in terms of performances obtained on heterogeneous platforms, in order to compare the different deployments of the \srl pattern.
The general methodology we adopt is to derive a \srl formulation of the considered
problem, translate it into a \ff pattern and compare different deployments of the \srl pattern.
Namely, we consider CPU, single-GPU and multi-GPU deployments. 
We remark, as we discussed in Sec.~\ref{sub:srl impl}, that the CPU deployment is a native multi-core implementation, thus not relying on OpenCL as parallel runtime.
Moreover, GPU deployments are compared to the best-case scenarios from the CPU world, thus considering the parallel configuration (e.g. thread allocation) of the \ff deployment yielding best performance.
Three applications are considered: the Helmholtz equation solver
based on iterative Jacobi method (Sec.~\ref{sub:helmholtz}), the Sobel edge
detector over image streams (Sec~\ref{sub:sobel}) and the two-phase video stream
restoration algorithm~\cite{ff:denoiser:ijhpca:15} (Sec.~\ref{sub:denoiser}).
All applications work on single-precision floating point data.
Each experiment was conducted on three different platforms: 
1) an Intel workstation with 2 eight-core
(2-way hyper-threading) Xeon E5-2660 @2.2GHz, 20MB L3 shared cache,
and 64 GBytes of main memory, equipped with two NVidia Tesla M2090 GPUs;
2) an Intel workstation with one
eight-core (2-way hyper-threading) Xeon E5-2650 @2.6GHz, 20MB L3 shared cache,
64 GBytes of main memory, equipped with a high-end NVidia Tesla K40 GPU;
3) a small Samsung workstation with a eight-core Exynos-5422 CPU 
(quad core Cortex-A15 @2.0GHz plus quad core Cortex-A7 @1.4 GHz)
equipped with a Arm Mali-T628 GPU. All systems run Linux x86\_64.


\subsection{The Helmholtz equation solver}
\label{sub:helmholtz}
The first application we consider is an iterative solver for the Helmholtz partial differential equation,
which is applied in the study of several physical problems. The solver is a paradigmatic
case of iterative 2D-stencil computation, in which each point of a read-only matrix (i.e. the input matrix)
is combined with the respective 3-by-3 neighbourhood of the partial solution matrix in order to compute a new partial solution.
The termination is based on a convergence criterion, evaluated as a function of the difference between two partial solutions at successive iterations, compared against a global threshold.
\begin{table}[t]
\textsf{\tsize
\begin{center}
\begin{tabular}{lrrrr}
\toprule
 Platform & Rows & CPU (\si{s}) & 1xGPU (\si{s}) & 2xGPUs 1:2 (\si{s}) \\ \midrule
\multirow{3}{*}{\parbox{3cm}{\scriptsize 2 eight-core Xeon @2.2GHz,\\
2 Tesla M2090 GPUs}} 
 & 512 & 0.31 & 0.31 & 0.32 \\
 & 4096 & 16.99 &10.84 & 5.88 \\
 & 16384 & 252.67 &171.84& 91.46 \\ \midrule
\multirow{3}{*}{\parbox{3cm}{\scriptsize 1 eight-core Xeon @2.6GHz,\\
Tesla K40 GPU}} 
 & 512 & 0.26 & 0.26 &- \\
 & 4096 & 25.00 & 7.42 &- \\
 & 16384 & 384.16 & 116.37 &- \\ \midrule
\multirow{3}{*}{\parbox{3cm}{\scriptsize Quad A15 @2.0GHz +\\
Quad A7 @1.4GHz,\\
Arm Mali-T628 GPU}} 
 & 512 & 3.51 & 6.91 &- \\
 & 2048 & 13.87 & 23.83 &- \\
 & 4096 & 64.61 & 92.51 &-\\
\bottomrule
\end{tabular}
\end{center}
}
\caption{Execution time of the Helmholtz equation solver.
	Convergence is reached after 10 iterations.\label{tab:helmholtz}}
\end{table}

The implemented \ff pattern is a single \srl pattern executing the procedure
over different input matrices. Table~\ref{tab:helmholtz}
shows the observed results.
The general behaviour, except for the third platform discussed later, is an immediate improvement resulting from the
GPU exploitation. A cross-platform exception is the small matrix case, on which the same execution times are observed on CPU and GPU deployments.
This is easily explained by communication overheads, as the ratio of H2D/D2H copies to actual computation is non-negligible in that case.
Speedups exhibited by the K40 and the M2090 GPUs mirror both the different computational capabilities of the two devices and the CPU parallelism available on the respective platforms.
Moreover, on the first platform execution times on the 1:2 two-GPU deployment scales almost linearly with respect to the one-GPU deployment. This shows that the multi-GPU runtime does not introduce any substantial overhead while managing data distribution and synchronising for halo-swap, when increasing the level of parallelisation in our implementation.
Finally, the third platform shows some inefficiency in this case, that could be addressed by providing careful optimisations tailored to this platform.

\subsection{The streaming Sobel edge detector}
\label{sub:sobel}

The second application we consider is a classical image processing filter,
namely the Sobel edge detector. It is a simple non-linear convolution-like operator, which
applies a 2D-stencil to each (3-by-3 neighbourhood of the) pixel of the input image to produce a new image,
in which pixel values represent the likelihood of the pixel belonging to an edge in the original image.
As with all the convolution-like image processing filters, the Sobel detector is a paradigmatic case of
non-iterative 2D-stencil computation.
The streaming variant applies the Sobel filter to a series of independent images, each from a different file.

We implemented a \srl version of the Sobel filter, which arises directly from its definition.
We applied the filter to three different square input images, with different sizes.
Moreover, we included a streaming version in order to both consider a more common use case
and show the approach of integrating a data-parallel pattern (the basic Sobel filter) into a \ff pattern.
The resulting pattern is: \textsf{pipe(read, sobel, write)}, where \textsf{sobel} is a \srl pattern
and \textsf{pipe(a,b)} is the classical pipeline with functional semantics $\textsf{b}\circ\textsf{a}$
and executing \textsf{a} and \textsf{b} in parallel over independent items.
We ran the streaming version on streams of 100 images, each built as random
permutation of the input set mentioned. Different deployments
have been compared over the same stream, kept constant by fixing the random seed.
Because of the reduced amount of GPU memory available on the third platform,
we excluded the largest image from tests.

\begin{table}[t]
\textsf{\tsize
\begin{center}
\begin{tabular}{lrrrr}
\toprule
 Platform & Width (px) & CPU (\si{s}) & 1xGPU (\si{s}) & 2xGPUs 1:2 (\si{s})\\
 \midrule
\multirow{4}{*}{\parbox{3cm}{\scriptsize 2 eight-core Xeon @2.2GHz,\\
2 Tesla M2090 GPUs}} 
 & 512 & 0.33 \si{\ms} & 0.79 \si{\ms} & 1.33 \si{\ms}\\
 & 4096 & 0.02 & 0.02 & 0.01\\
 & 16384 & 0.22 & 0.31 & 0.20\\
 & Stream & 11.96 & 16.27 & 11.09\\
 \midrule
\multirow{4}{*}{\parbox{3cm}{\scriptsize 1 eight-core Xeon @2.6GHz,\\
Tesla K40 GPU}} 
 & 512 & 0.58 \si{\ms} & 0.68 \si{\ms} &-\\
 & 4096 & 0.03 & 0.01 &-\\
 & 16384 & 0.53 & 0.17 &-\\
 & Stream & 27.89 & 8.97&-\\
 \midrule
\multirow{3}{*}{\parbox{3cm}{\scriptsize Quad A15 @2.0GHz +\\
Quad A7 @1.4GHz,\\
Arm Mali-T628 GPU}} 
 & 512 & 4.91 \si{\ms} & 7.02 \si{\ms} &-\\
 & 4096 & 0.29 & 0.27 &-\\
 & Stream & 28.22 & 23.45 &-\\
\bottomrule
\end{tabular}
\end{center}
}
\caption{Execution time of the Sobel filter on different platforms.
For each platform, the upper rows refer to the single-item cases (i.e. restoration of single pictures);
the last row refers to the streaming variant on 100 random images.\label{tab:sobel}}
\end{table}

Table~\ref{tab:sobel} shows the observed results.
We remark that the single-iteration pattern represents the worst-case scenario for
GPU exploitation, since little computation is available to hide the latency of H2D/D2H memory copies.
Indeed, the CPU deployment on the first platform performs better than the single-GPU one,
while the 1:2 two-GPU deployment still yields some improvement.
Conversely, the K40 GPU on the second platform is still able to improve the execution time
by an average of about $3\times$ with respect to the CPU deployment.
Finally, small improvement is obtained by the Mali GPU on the third platform, while a more substantial
improvement is observable in the streaming variant, since in the latter case the GPU-side allocation overhead 
is mitigated.



\subsection{The two-phase video restoration algorithm}
\label{sub:denoiser}

The third and most complex application is a two-phase parallel video restoration filter.
For each video frame, in the first step (i.e. the detection phase) a traditional adaptive median filter is employed for detecting noisy pixels,
while in the second step (i.e. the restoration phase) a regularisation procedure is iterated until the noisy pixels are replaced
with values which are able to preserve image edges and details. The restoration phase is based on a 2D-stencil regularisation procedure,
which replaces each pixel with the value minimising a function of the pixel neighbourhood.
The termination is decided on a simple convergence criterion, based on the average absolute difference between
two partial solutions at successive iterations, compared against a global threshold.

We implemented the application
by modelling it with the \ff pattern:
\textsf{{pipe(read, detect, ofarm(restore), write)}},
where \textsf{restore} is the \srl implementation of the restoration procedure
and \textsf{ofarm(a)} is a pattern in which input items
are processed in parallel by multiple instances of the \textsf{a} pattern and the order is preserved
in the output stream.
Samples of 100 frames at VGA ($640 \times 480$), 720p ($1280 \times 720$)
and HDTV ($2048 \times 1080$) resolutions are considered as input streams and artificial noise is added
to each stream, at $30\%$ and $70\%$ level.
In order to include an example of different integration schemata of a \srl pattern into a \ff pattern,
both 1:1 and 1:2 deployments are considered.

\begin{table}[t]
\textsf{\tsize
\begin{center}
\begin{tabular}{lcrrrr}
\toprule
Platform & Video & CPU (\si{s}) & 1xGPU (\si{s}) & 2xGPUs 1:1 (\si{s}) & 2xGPUs 1:2 (\si{s})\\
 \midrule
\multirow{6}{*}{\parbox{2.5cm}{\scriptsize 2 eight-core\\Xeon @2.2GHz,\\
2 Tesla\\M2090 GPUs}} 
 & VGA, 30\% &23.74 &8.69&4.59&4.64\\
 & VGA, 70\% & 49.65&8.70&4.61&4.69\\
 & 720p, 30\% & 67.78&25.23&13.12&13.16\\
 & 720p, 70\% & 147.69&25.28&13.50&13.55\\
 & 1080p, 30\% & 162.27&60.01&30.78&30.81\\
 & 1080p, 70\% &354.18 &60.11&32.39&32.44\\ \midrule
\multirow{6}{*}{\parbox{2.5cm}{\scriptsize 1 eight-core\\Xeon @2.6GHz,\\
Tesla K40 GPU}} 
 & VGA, 30\% & 41.56 & 3.41&-&-\\
 & VGA, 70\% &{87.32} &{4.39}&-&-\\
 & 720p, 30\% &{118.99} &{9.72}&-&-\\
 & 720p, 70\% &{259.54} &{12.71}&-&-\\
 & 1080p, 30\% &{285.34} &{23.89}&-&-\\
 & 1080p, 70\% &{623.20} &{29.99}&-&-\\ \midrule
\multirow{6}{*}{\parbox{2.5cm}{\scriptsize Quad A15 @2.0GHz+\\
Quad A7 @1.4GHz,\\
Arm Mali-T628 GPU}} 
 & VGA, 30\% & 373.63 & 144.57 &-&-\\
 & VGA, 70\% & 739.92 & 206.26 &-&-\\
 & 720p, 30\% & 986.55 & 409.77 &-&-\\
 & 720p, 70\% & 2125.89 & 601.42 &-&-\\
 & 1080p, 30\% & 2730.52 & 974.87 &-&-\\
 & 1080p, 70\% & 4644.86 & 1364.74 &-&-\\ \bottomrule
\end{tabular}
\end{center}
}
\caption{Execution time of the restoration filter over
	100-frame video samples.\label{tab:exp-denoiser}}
\end{table}

Table~\ref{tab:exp-denoiser} shows the observed results.
As expected, the multi-iteration streaming nature exhibited by this application
is profitably captured by the \srl pattern.
Both the reuse of device memory across different input items and
the considerable amount of computation per iteration exhibited by this application
(convergence is reached in 10 to 30 iterations)
yield good performance in all of the scenarios considered.
In particular, execution times on the K40 GPU on the second platform show speedups ranging
from $12\times$ to $20\times$ with respect to the CPU deployment,
delivering a throughput of about 30 frames per second for the low-noise case on VGA resolution.
Analogous performances are obtained from the 1:1 two-GPU deployment on the second platform,
while a minimal degradation is introduced by switching to the 1:2 deployment, due to the
slightly higher number of synchronisations induced as discussed in \ref{sub:srl impl}.
Also, the third platform provides considerable speedup in this case,
confirming that it is well suited to target media-oriented applications,
which do not feature high numerical demand.